\documentclass[aps,prd,12point,twocolumn,nofootinbib,showpacs,superscriptaddress]{revtex4-1}
\def\theequation{\arabic{section}.\arabic{equation}}
\usepackage{psfrag}
\usepackage{subfigure}
\usepackage{color}
\usepackage{mathrsfs}
\usepackage{graphicx}
\usepackage{amssymb, bm}
\usepackage{amsmath, amsthm}
\usepackage{epstopdf}
\usepackage{hyperref}
\usepackage{enumerate}
\usepackage{longtable}

\newcommand{\be}{\begin{equation}}
\newcommand{\ee}{\end{equation}}
\def\Box{\hbox{$\rlap{$\sqcup$}\sqcap$}}

\begin{document}
\def\theequation{\arabic{section}.\arabic{equation}}


\title{Wyman's ``other'' scalar field solution, Sultana's generalization,
and their Brans-Dicke and ${\cal R}^2$ relatives}


\author{Ali Banijamali}
\email[]{a.banijamali@nit.ac.ir}
\altaffiliation{}
\affiliation{Department of Basic Sciences, Babol
Noshirvani University of Technology, Babol, Iran}
\affiliation{Department of Physics \& Astronomy, Bishop's University\\
2600 College Street, Sherbrooke, Qu\'ebec, Canada J1M~1Z7
}

\author{Behnaz Fazlpour}
\email[]{b.fazlpour@umz.ac.ir}
\affiliation{Department of Physics, Babol Branch, Islamic Azad
University, Babol, Iran}
\affiliation{Department of Physics \& Astronomy, Bishop's University\\
2600 College Street, Sherbrooke, Qu\'ebec, Canada J1M~1Z7
}

\author{Valerio Faraoni}
\email[]{vfaraoni@ubishops.ca}
\affiliation{Department of Physics \& Astronomy, Bishop's University\\
2600 College Street, Sherbrooke, Qu\'ebec, Canada J1M~1Z7
}


\begin{abstract}

Wyman's less known static and spherically symmetric solution of the
Einstein-Klein-Gordon equations and its recent generalization for positive
cosmological constant are discussed, showing that they contain central
naked singularities.  By mapping back to the 
Jordan frame, we obtain the
conformal cousins of these geometries that solve the vacuum Brans-Dicke
field equations and are time-dependent and spherical. Their physical
nature is discussed and the geometry is shown to be also a solution of
purely quadratic gravity.

\end{abstract}


\maketitle

\section{Introduction}
\label{sec:1}
\setcounter{equation}{0}

Scalar fields are ubiquitous in theoretical physics, beginning with
particle physics, where one encounters the Higgs field of the Standard
Model \cite{Higgs1, Higgs2} and other scalar fields in its supersymmetric
extensions; in gravitation, beginning with the gravitational
scalar of Brans-Dicke theory \cite{BD} and its scalar-tensor extensions
\cite{ST}, the dilaton of string theories \cite{GreenSchwarzWitten,
Polchinski}, and the moduli fields of multi-dimensional physics. The
low-energy limit of the bosonic string theory produces an $\omega=-1$
Brans-Dicke theory \cite{Callanetal85, FradkinTseytlin85}.

The general solution of the Einstein
equations sourced by a minimally
coupled, free and massless scalar field\footnote{The reason for using
a tilde over the  quantities appearing in this section will
be clear in the following, where the same variables are
regarded as Einstein frame variables of a scalar-tensor
theory, which are usually denoted with a tilde.} $\tilde{\phi}$ that
is static, spherically symmetric, and asymptotically flat is well known
and has been rediscovered several times. Under the assumption that
the matter field $\phi$ depends only on the radial coordinate, the
unique solution was found by Fisher \cite{Fisher48} and, in other
coordinates or in other forms, by Bergmann and Leipnik \cite{BL57},
Janis, Newman and Winicour \cite{JNW68}, Buchdahl \cite{Buchdahl72}, and
Wyman \cite{Wyman81} (``FBLJNWBW''). Wyman gave the most general form of
the
solution. In doing so, he showed that another family of solutions
can be obtained by assuming the geometry to be spherically symmetric
and static, while the scalar depends only on time,
$\tilde{\phi}=\tilde{\phi}(t)$. Unfortunately, this class of
solutions can, in general, be expressed only in the form of power
series, which is not useful from the calculational point of view.
However, one of the solutions has the almost trivial form
\begin{eqnarray}\label{act}
d\tilde{s}^{2}=-\kappa r^{2}dt^{2}+2dr^{2}+r^{2}d\Omega^{2}_{(2)},
\end{eqnarray}
\begin{eqnarray}\label{phi}
\tilde{\phi}(t)=\tilde{\phi}_{0}t,
\end{eqnarray}
where $\kappa= 8\pi G$, $G$ is Newton's constant and
$d\Omega^{2}_{(2)}=d\theta^{2}+\sin^{2}\theta \, d\varphi^{2}$ is the
line element on the unit two-sphere. We will refer to this solution
of the coupled Einstein-massless Klein-Gordon equations
\begin{eqnarray}
\tilde{\mathcal{R}}_{ab}-\frac{1}{2} \, \tilde{g}_{ab}\tilde{\mathcal{R}}
=\kappa\Big(\tilde{\nabla}_{a} \tilde{\phi}
\tilde{\nabla}_{b}
\tilde{\phi} -\frac{1}{2}\tilde{g}_{ab}\tilde{\nabla}^{c} \tilde{\phi}
\tilde{\nabla}_{c} \tilde{\phi} \Big) \,,
\end{eqnarray}
\begin{eqnarray}
\tilde{\Box} \tilde{\phi} =0 \,,
\end{eqnarray}
(where $\tilde{\Box} \equiv \tilde{g}^{ab} \tilde{\nabla}_a
\tilde{\nabla}_b$ is the d'Alembertian) as Wyman's ``other'' solution.
Recently, this solution has been
generalized by Sultana \cite{Sultana15} to include a positive
cosmological constant $\Lambda$. The geometry changes according to
\begin{eqnarray}\label{geo}
d\tilde{s}^{2}=-\kappa r^{2}dt^{2}+\frac{2dr^{2}}{1-\frac{2\Lambda
r^{2}}{3}}+r^{2}d\Omega^{2}_{(2)},
\end{eqnarray}
while the scalar field is still linear in time and given by
Eq.~(\ref{phi}). Wyman's "other" solution is recovered, of course, in
the limit
$\Lambda\rightarrow 0$.

In this article we clarify the physical nature of the Sultana and
the Wyman solutions, which was left unclear in the literature thus
far, and we regard Sultana's solution as the Einstein frame version
of a solution of Brans-Dicke theory. In Sec.~\ref{sec:3}, by mapping 
back this
solution\footnote{In
Ref.~\cite{Sultana15}, Sultana generated new solutions of
conformally coupled scalar field theory with a Higgs potential using
a technique developed by Bekenstein \cite{Bekenstein}.}
to the Jordan frame,  we generate a new 
two-parameter class of
solutions of the Brans-Dicke field equations.

In general, the mapping from the Einstein to the Jordan frame generates
solutions with very contrived Jordan frame potentials, even though the
Einstein frame
potential is physically well motivated. However, in our case the constant
Einstein
frame potential $U=\Lambda/\kappa$ of Sultana's solution,
when mapped to the Jordan conformal frame, generates a physically
relevant mass term $V(\phi)=m^2 \phi^{2}/2 $ for the Jordan
frame scalar $\phi$.  The new Brans-Dicke solution is derived in 
Sec.~\ref{sec:3}, while Sec.~\ref{sec:4} shows that this geometry solves 
also the field equations of quadratic $f({\cal R})$ gravity. 
Sec.~\ref{sec:5} contains the conclusions.  We follow the notation of Ref. 
\cite{Waldbook}; the metric signature
is $-+++$, and we use units in which Newton's constant $G$ and the
speed of light $c$ are unity.

\section{Sultana solution and its Wyman limit}
\label{sec:2}
\setcounter{equation}{0}

The most general spherically
symmetric line element can be written in the form
\begin{eqnarray}
d\tilde{s}^{2}=-A^{2}(t,r)dt^{2}+B^{2}(t,r)
dr^{2}+r^{2}d\Omega^{2}_{(2)} \,.
\end{eqnarray}
Moreover, if only the cosmological constant $\Lambda$ is present, the
scalar field satisfies
\begin{eqnarray}\label{box}
\tilde{\Box}\tilde{\phi}=0 \,,
\end{eqnarray}
since the constant potential $U(\tilde{\phi})= \Lambda/\kappa $ gives
zero contribution to the full Klein-Gordon equation
$\tilde{\Box} \tilde{\phi} - dU /d\tilde{\phi}=0$. The coupled
Einstein-Klein-Gordon equations assume the form
\begin{widetext}
\begin{eqnarray}\label{B}
\frac{\dot{B}}{B}=4\pi r\dot{ \tilde{\phi}} \tilde{\phi}' \,,
\end{eqnarray}
\begin{eqnarray}
\frac{2B'}{B^{3}r}-\frac{1}{B^{2}r^{2}}+\frac{1}{r^{2}}
=\frac{\kappa}{A^{2}}\Big[\dot{  \tilde{\phi}}^{2}
+\frac{A^{2}}{2}\Big(-\frac{\dot{  \tilde{\phi}}^{2}}{A^{2}}
+\frac{  \tilde{\phi}'^{2}}{B^{2}}\Big)+\frac{\Lambda A^{2}}{\kappa}\Big]
\,,
\end{eqnarray}

\begin{eqnarray}
\frac{2A'}{Ar}-\frac{B^{2}}{r^{2}}+\frac{1}{r^{2}}
= \kappa\Big[
\tilde{\phi}'^{2}-\frac{B^{2}}{2}\Big(-\frac{\dot{
\tilde{\phi}}^{2}}{A^{2}}
+\frac{ \tilde{\phi}'^{2}}{B^{2}}\Big)-\frac{\Lambda B^{2}}{\kappa}\Big]
\,,
\end{eqnarray}

\begin{eqnarray}\label{Ein}
\frac{r}{B^{3}}\Big(\frac{A' B}{A}-B'-\frac{r
B^{2}\ddot{B}}{A^{2}}+\frac{r \dot{A}B^{2}\dot{B}}{A^{2}}-\frac{r A'
B'}{A}+\frac{rA''B}{A}\Big)
=\kappa\Big[-\frac{r^{2}}{2}\Big(-\frac{\dot{ \tilde{\phi}}^{2}}{A^{2}}
+\frac{ \tilde{\phi}'^{2}}{B^{2}}\Big)-\frac{\Lambda r^{2}}{\kappa}\Big]
\,,
\end{eqnarray}
\end{widetext}
where an overdot and a prime denote differentiation with respect to $t$
and $r$, respectively. It is straightforward to verify that (\ref{geo}),
(\ref{phi})
satisfies Eqs.~(\ref{B})-(\ref{Ein}) and (\ref{box}) provided that
$\tilde{\phi}_{0}=1$ (in
the notation adopted, $t$ is a dimensionless coordinate).

Let us discuss the Sultana solution (\ref{geo}), (\ref{phi}) with
regard to its physical properties. The coordinates $t$ and $r$ span
the ranges
\begin{eqnarray}
-\infty<t<+\infty,\,\,\,\,\,\,\,\, \;\; 0\leq r<\sqrt{\frac{3}{2\Lambda}}.
\end{eqnarray}
As will be clear shortly, the presence of $\Lambda>0$ causes the
presence of a cosmological-type horizon at
$r_{H}=\sqrt{\frac{3}{2\Lambda}}$ and the locally static coordinates
$\left( t,r,\theta,\varphi \right)$ cover only the patch below this
horizon.

In the limit $\Lambda\rightarrow 0$, Wyman's ``other'' solution
(\ref{act}), (\ref{phi}) is recovered, the horizon is pushed to
infinity, and $ 0\leq r\leq +\infty$. Interestingly, the Wyman geometry
(\ref{act}) corresponds to a special case of a geometry found by
Carloni and Dunsby \cite{CarloniDunsby} in scalar-tensor gravity,
but with a scalar $\tilde{\phi}=\tilde{\phi}(r)$ that depends
only on the radial (instead of the
temporal) coordinate and for a power-law potential.

The Ricci scalar of the Sultana solution is
\begin{eqnarray}
 \tilde{\mathcal{R}}=\kappa \, \tilde{g}^{ab}\,\tilde{\nabla}_{a}
\tilde{\phi}\,
\tilde{\nabla}_{b}\tilde{\phi}
=4\Lambda-\frac{\tilde{\phi_{0}}^{2}}{r^{2}} \,, \end{eqnarray}
and it diverges as $r\rightarrow 0^{+}$, where the coordinate $r$
coincides with the physical (areal) radius. If present, horizons are
located by the roots of the equation
\begin{eqnarray}
\tilde{g}^{ab}\,\tilde{\nabla}_{a}r\tilde{\nabla}_{b}r=\tilde{g}^{rr}=\Big(1-\frac{2\Lambda
r^{2}}{3}\Big)\frac{1}{2}=0 \,,
\end{eqnarray}
which provides the radius of the cosmological horizon
\begin{eqnarray}
r_{H}=\sqrt{\frac{3}{2\Lambda}}
\end{eqnarray}
created by $\Lambda$. This horizon is smaller than the de Sitter
horizon of radius $\sqrt{3/\Lambda}$. There are no other
horizons, therefore
the Sultana solution is interpreted as a naked central singularity
embedded in a ``background'' created by the cosmological constant.
(We use quotation marks because, due to the non-linearity of the 
field equations, it is impossible to split a metric into a 
``background'' and a ``deviation'' from it in a covariant way (barring 
special algebraic structures as, {\em e.g.}, in generalized Kerr-Schild 
metrics, but this is not the case here.)

In the $\Lambda\rightarrow 0$ limit corresponding to Wyman's ``other''
solution, there is no horizon and the 
geometry can be interpreted physically as a naked 
singularity embedded in a space which is not asymptotically 
flat since, in this case, the Ricci tensor
\be
\tilde{{\cal R}}_{ab}= \kappa \nabla_a \tilde{\phi} \nabla_b \tilde{\phi} 
= \kappa \tilde{\phi}_0^2 \delta_{a0} \delta_{b0} 
\ee
does not depend on $r$ and does not go to zero as $r\rightarrow 
+\infty$.

The Misner-Sharp-Hernandez mass \cite{MSH} contained in a ball of radius
$r$ is given
by $ 1-2M_{MSH}/r \equiv \nabla^{c} r \nabla_{c}r$, which yields
\begin{equation}
M_{MSH}(r)=\frac{r}{4}\Big(1+\frac{2\Lambda r^{2}}{3}\Big) \,.
\end{equation}
Comparing with the Misner-Sharp-Hernandez mass of de Sitter space, which
is $M_{dS}= \Lambda r^{3}/6 $, we see that the scalar field gives
an additional positive contribution $ r/4 $.

\section{The Brans-Dicke counterpart of Sultana's solution}
\label{sec:3}
\setcounter{equation}{0}

Let us consider vacuum (Jordan frame) Brans-Dicke theory, described
by the action
\begin{eqnarray}\label{action}
S_{BD}= \int d^{4}x \,
\frac{\sqrt{-g}}{8\pi}\Big[\phi\mathcal{R}-\frac{\omega}{\phi}
\, \nabla^{c}\phi\nabla_{c}\phi-V(\phi)\Big]
\end{eqnarray}
where $\phi>0$ is the Brans-Dicke field, the constant $\omega$ is
the ``Brans-Dicke coupling", and $g$ is the determinant of the
metric.  The conformal transformation of the metric
\begin{eqnarray}\label{met}
g_{ab}\rightarrow\tilde{g}_{ab}=\Omega^{2}g_{ab}=\phi \, g_{ab} \,,
\end{eqnarray}
and the scalar field redefinition
\begin{eqnarray}\label{trans}
\phi\rightarrow\tilde{\phi}=
\sqrt{\frac{|2\omega+3|}{16\pi}}\ln\Big(\frac{\phi}{\phi_{\ast}}\Big)
\end{eqnarray}
(where $\phi_{\ast}$ is a positive constant) recast the Brans-Dicke
action in its Einstein frame form
\begin{eqnarray}
S_{BD}&=&\int d^{4}x \sqrt{-\tilde{g}}  \,
\Big[\frac{\tilde{\mathcal{R}}}{16\pi}-
\frac{1}{2} \, \tilde{g}^{ab}\,\tilde{\nabla}_{a}\tilde{\phi}\,
\tilde{\nabla}_{b}\tilde{\phi}-U(\tilde{\phi})\Big] \,,\nonumber\\
&&
\end{eqnarray}
where
\begin{eqnarray}\label{U}
U(\tilde{\phi})=\frac{V(\phi)}{\phi^{2}}\Big|_{\phi=\phi(\tilde{\phi})}
\end{eqnarray}
and a tilde denotes Einstein frame quantities. In this form, the
theory has all the appearances of general relativity with a
minimally coupled scalar field as a source. Since the Sultana
solution solves the corresponding field equations with constant
potential $U(\tilde{\phi})= \Lambda / \kappa$, we can map this
solution to the Jordan frame to obtain a new family of solutions of
the vacuum Brans-Dicke field equations. The potential ruling the dynamics
of the gravitational Brans-Dicke field $\phi$ in the Jordan frame is
given by Eq.~(\ref{U}) as the simple mass term
\begin{eqnarray}
V(\phi)=\frac{m^{2}\phi^{2}}{2}
\end{eqnarray}
where
\begin{eqnarray}
m^{2}=\frac{2\Lambda}{\kappa}>0 \,.
\end{eqnarray}
The Jordan frame potential $V(\phi)$ no longer contains a cosmological
constant and the original parameter $\Lambda$ now parametrizes the scalar
field mass $m$.
The solution-generating technique used here, originally introduced by
Bekenstein for non-minimally coupled scalar fields (a different
scalar-tensor theory), has been
employed a number of times in the literature on scalar-tensor
gravity. By applying Eqs. (\ref{met}) and (\ref{trans}) to the
Sultana solution (\ref{geo}), (\ref{phi}) one finds
\begin{eqnarray}
ds^{2}&=&\phi^{-1} \, d\tilde{s}^{2}=
\mbox{e}^{-\sqrt{\frac{16\pi}{|2\omega+3|}}
\, \tilde{\phi_{0}}t} \nonumber\\
&&\nonumber\\
& \, & \cdot \Big(-\kappa
r^{2}dt^{2}+\frac{2dr^{2}}{1-\frac{2\Lambda
r^{2}}{3}}+r^{2}d\Omega^{2}_{(2)}\Big) \,,\\
&&\nonumber\\
\phi(t)&=&\phi_{\ast}
\mbox{e}^{\sqrt{\frac{16\pi}{|2\omega+3|}} \, \tilde{\phi_{0}}t}.
\end{eqnarray}
Since $\phi_{\ast}>0$, the Jordan frame scalar is positive. It is
convenient to introduce the new time coordinate $\tau$ defined by
$$d\tau=e^{-\sqrt{\frac{4\pi}{|2\omega+3|}}\tilde{\phi_{0}}t}dt,$$
or
\begin{eqnarray}
\tau=
\sqrt{\frac{|2\omega+3|}{4\pi\tilde{\phi_{0}}^{2}}}
\bigg[1-\mbox{e}^{-\sqrt{\frac{4\pi}{|2\omega+3|}} \,
\tilde{\phi_{0}}t}\bigg] \,,
\end{eqnarray}
by imposing that $\tau(t=0)=0$. Introducing
\begin{eqnarray}
\tau_{\ast}\equiv\sqrt{\frac{|2\omega+3|}{4\pi\tilde{\phi_{0}}^{2}}} \,,
\end{eqnarray}
the new Brans-Dicke solution reads
\begin{eqnarray}
ds^{2} &=& -\kappa
r^{2}d\tau^{2}+\Big(1-\frac{\tau}{\tau_{\ast}}\Big)^{2}
\Big(\frac{2dr^{2}}{1-\frac{2\Lambda
r^{2}}{3}}+r^{2}d\Omega^{2}_{(2)}\Big) \,, \nonumber\\
&& \label{new1}\\
\phi(\tau) &=&  \frac{\phi_{\ast}}{\Big(1-\frac{\tau}{\tau_{\ast}}
\Big)^{2}}  \,. \label{new2}
\end{eqnarray}
In this solution, $\omega$ and $\Lambda$ are parameters of the
theory, while $\phi_{\ast}$ plays the role of an initial condition.

\subsection{Nature of the solution}

Let us examine the nature of this new solution. The Ricci scalar is
\begin{eqnarray}
\mathcal{R} &=& \frac{\omega}{\phi^{2}}\nabla^{c}\phi
\nabla_{c}\phi+\frac{3\Box\phi}{\phi}+\frac{2V}{\phi} \nonumber\\
&&\nonumber\\
& = & \frac{1}{\kappa\Big(1-\frac{\tau}{\tau_{\ast}}\Big)^{2}}
\Big(2\Lambda\phi_{\ast}-\frac{4\omega}{\tau_{\ast}^{2}r^{2}}\Big) \,.
\label{Ricci}
\end{eqnarray}
For any value of $\omega$, the Ricci scalar diverges as
$\tau\rightarrow\tau_{\ast}^{-}$; this value of the time coordinate
corresponds to a Big Crunch-type singularity, where the scalar
$\phi$ also diverges.

If $\omega\neq0$, $\mathcal{R}$ diverges also as $r\rightarrow0^{+}$
(the case $\omega=0$ will be analyzed separately). The areal radius
is
\begin{eqnarray}
R(\tau,r)=\Big(1-\frac{\tau}{\tau_{\ast}}\Big)r \,,
\end{eqnarray}
and $R\rightarrow 0$ as $r\rightarrow 0$. Hence, there is a central
singularity if $\omega\neq0$. Next, one would like to answer the question
of whether this singularity is covered by horizons. If present,
horizons are located by the roots of
\begin{eqnarray}
\nabla^{c}R\nabla_{c}R
&=& g^{00}\Big(\frac{dR}{d\tau}\Big)^{2}+g^{11}\Big(\frac{dR}{dr}\Big)^{2}
\nonumber\\
&&\nonumber\\
&=&\frac{1}{2} \Big[\frac{-2}{\kappa\tau_{\ast}^{2}}+1-\frac{2\Lambda
r^{2}}{3}\Big] \,.
\end{eqnarray}
As in the case of the Sultana spacetime,
this is a horizon created by the cosmological constant
at
\begin{eqnarray}
r_{H}&=& \sqrt{\frac{3}{2\Lambda}}
\sqrt{1-\frac{8\pi\tilde{\phi_{0}}^{2}}{|2\omega+3|\kappa}}
=\sqrt{\frac{3}{2\Lambda}} \sqrt{1-\frac{2}{\kappa\tau_{\ast}^{2}}}
\,.\nonumber\\
&&
\end{eqnarray}
We have again a naked central singularity embedded in a ``background''
created by $\Lambda$ and $\phi$, which ends its history at a finite
future $\tau_{\ast}$.

\subsection{The case $\omega=0$}

For completeness, we now discuss the case $\omega=0$ in which the
Ricci scalar (\ref{Ricci}) is regular as $r\rightarrow 0^{+}$. In this
case, the square of the Ricci tensor
$\mathcal{R}_{ab}=\phi^{-1}\Big(\nabla_{a}\nabla_{b}\phi+Vg_{ab}/2\Big)$
is
\begin{eqnarray}
\mathcal{R}_{ab}\mathcal{R}^{ab}
=\frac{1}{\phi^{2}}\Big(\nabla_{a}\nabla_{b}\phi\nabla^{a}\nabla^{b}\phi
+\frac{\Lambda^{2}}{\kappa^{2}}\Big) \,.
\end{eqnarray}
Since
\begin{equation}
\nabla_{a}\nabla_{b}\phi=\partial_{a}\partial_{b}\phi-\Gamma_{ab}^{c}
\partial_{c}\phi \,,
\end{equation}
\begin{equation}
\partial_{a}\partial_{b}\phi=\frac{6\phi_{\ast}}{\tau_{\ast}^{2}
\Big(1-\frac{\tau}{\tau_{\ast}}\Big)^{4}}\delta_{a0}\delta_{b0} \,,
\end{equation}
and
\begin{eqnarray}
\Gamma_{ab}^{c}\partial_{c}\phi=\frac{4\phi_{\ast}}{r\Big(1-
\frac{\tau}{\tau_{\ast}}\Big)^{3}\tau_{\ast}}\Big(\delta_{a0}\delta_{b1}
+\delta_{a1}\delta_{b0}\Big) \,,
\end{eqnarray}
one obtains after straightforward manipulations
\begin{eqnarray}
\mathcal{R}_{ab}\mathcal{R}^{ab} &=& \frac{1}{\tau_{\ast}^{4}\kappa
r^{4}\Big(1-\frac{\tau}{\tau_{\ast}}\Big)^{4}}
\Big(\frac{9}{\kappa\tau_{\ast}^{2}}-4+\frac{8\Lambda
r^{2}}{3}\Big) \nonumber\\
&&\nonumber\\
&\, & +\frac{\Lambda^{2}}{\kappa^{2}\phi_{\ast}^{2}}\Big(1
-\frac{\tau}{\tau_{\ast}}\Big)^{4}\,. \label{boh}
\end{eqnarray}
This quantity diverges as $r\rightarrow 0^{+}$ and $R\rightarrow
0^{+}$, therefore the naked singularity at $R=0$ persists for
$\omega=0$.

In the special case $\Lambda=0$, we have the Brans-Dicke counterpart of
Wyman's ``other'' solution for which $\mathcal{R}=0$ but
$\mathcal{R}_{ab}\mathcal{R}^{ab}\rightarrow\infty$ as $R\rightarrow
0^{+}$.

\section{A new solution of purely quadratic gravity}
\label{sec:4}
\setcounter{equation}{0}

It is well known that metric $f({\cal R})$ gravity is equivalent to an
$\omega=0$ Brans-Dicke theory with a complicated potential \cite{review1,
review2, review3}. Specifically, the action
\be
S= \int d^4 x\, \frac{\sqrt{-g}}{16\pi} \, f({\cal R})
\ee
where $f$ is a non-linear function, can be shown \cite{review1,review2, review3} to be
dynamically equivalent
to the other action
\be
S= \int d^4 x\, \frac{\sqrt{-g}}{16\pi}  \left[ \phi {\cal R} -V(\phi)
\right] \label{questa}
\ee
if $f''\neq 0$, where $\phi=f'({\cal R})$ and
\be
V(\phi) = \phi {\cal R} - f({\cal R})\left|_{\phi=\phi( {\cal R})} \right.
\label{thispotential}
\ee
is a potential which, in general, is only defined implicitly. The
action~(\ref{questa}) is clearly an $\omega=0$ Brans-Dicke action with
this complicated potential for the scalar $\phi$. The question arises of
whether the geometry and scalar field obtained by setting $\omega=0$ in
Eqs.~(\ref{new1}) and~(\ref{new2}) provides a new solution of $f({\cal
R})$ gravity. If this is the case, then what is the corresponding form of
the function $f({\cal R})$?

To answer, one notes that Eqs.~(\ref{Ricci}) and~(\ref{new2}) give, for
$\omega=0$,
\be
{\cal R}= \frac{2\Lambda \phi_*}{\kappa \left( 1-\tau/\tau_* \right)^2}=
\frac{2\Lambda}{\kappa} \, \phi \,.
\ee
Then it must be
\be
\phi=f'({\cal R}) =\frac{\kappa {\cal R}}{2\Lambda} \,,
\ee
which integrates to
\be
f( {\cal R})= \frac{\kappa {\cal R}^2}{4\Lambda} +f_0 \,,
\ee
where $f_0$ is an integration constant corresponding to a new ({\em i.e.}, 
distinct
from $\Lambda$) cosmological constant in the theory. Now, by imposing that the
scalar field potential be given by Eq.~(\ref{thispotential}), one must
necessarily have
\be
V(\phi)= \frac{ \kappa {\cal R}^2}{4\Lambda} -f_0 \,.\label{ddelta1}
\ee
However, the condition $V(\phi)=m^2\phi^2/2$ must be satisfied
simultaneously, with the extra information that $m^2 =2\Lambda/\kappa$,
which yields
\be
V(\phi)= \frac{\kappa {\cal R}^2}{4\Lambda} \,.\label{ddelta2}
\ee
The comparison of Eqs.~(\ref{ddelta1}) and (\ref{ddelta2}) yields $f_0=0$.
Therefore, the time-dependent geometry~(\ref{new1}) (with $\omega=0$ in
the expression
of $\tau_*$) is a  solution of the fourth order  theory
\be
f({\cal R})=\frac{\kappa {\cal R}^2}{4\Lambda} \,.
\ee
This theory, which has scale-invariance properties, does not admit a
Newtonian limit \cite{PechlanerSexl66} and therefore is not adequate to
describe the present universe, but is instead a good approximation of the
Starobinski model $f({\cal R})= {\cal R} +\alpha
{\cal R}^2$ describing a viable (and even preferred by
observations \cite{Starob-obs}) scenario of inflation  \cite{Starobinsky}.
It is also the subject of several studies in the context of early universe
and of black hole physics \cite{rsquared}.

As already established following Eq.~(\ref{boh}), this spacetime hosts a naked central
singularity where, however, the Brans-Dicke scalar $\phi=f'({\cal R}) \propto {\cal R}$ is
regular. In addition, a spacelike singularity is reached simultaneously at all spatial points as
$\tau \rightarrow \tau_*$.

\section{Conclusions}
\label{sec:5}
\setcounter{equation}{0}

Scalar-tensor theories of gravity are the prototypical alternatives to
Einstein's general relativity and, in various forms (but especially
$f({\cal R})$
gravity, which is an $\omega=0$ Brans-Dicke theory with a complicated
potential \cite{review1, review2, review3}), they are studied intensively
in
order to explain the present acceleration of the universe without an {\em
ad hoc} dark energy \cite{FujiiMaeda, mymonograph, cosmography}. However, 
the
catalog of exact solutions of the scalar-tensor field equations, which
would allow a better understanding of these theories, is still very
limited. With this perspective, we have clarified in 
Sec.~\ref{sec:2} the physical nature of
Wyman's ``other'' solution \cite{Wyman81} and of its generalization
obtained by Sultana for a
 positive cosmological constant $\Lambda$
\cite{Sultana15}. These solutions describe  naked 
central singularities embedded  in curved ``backgrounds'' created by 
$\phi$, or by 
$\Lambda$ and $\phi$, respectively.  For comparison, the more well known 
FBLJNWBW solution
\cite{Fisher48, BL57, JNW68, Buchdahl72, Wyman81} and its Brans-Dicke
counterpart always describe naked singularities or wormhole throats
\cite{VFAT}. The solutions of the Einstein equations discussed here 
do not violate the Birkhoff theorem because: a)~they are not vacuum 
solutions, and b)~they are not asymptotically flat.

While the FBLJNWBW solution and its Brans-Dicke counterpart are static,
our new Brans-Dicke solution derived in Sec.~\ref{sec:3} is 
time-dependent and inhomogeneous. Only a
handful of such time-dependent solutions, which are harder to obtain, are
presently known in Brans-Dicke theory \cite{climobarr, myclimobarr,
confonarev} or $f({\cal R})$ gravity \cite{Clifton, myclifton,
confonarev}.
Further, the new solution corresponds to a mass term potential, which is 
physically very
reasonable and even desired, while the solution-generating technique that
we employed ({\em i.e.}, mapping from the Einstein to the Jordan frame)
usually produces contrived and unphysical scalar field potentials
$V(\phi)$ in the
Jordan frame. Realistic scalar field configurations seem to collapse to
black holes without hair \cite{nohair}, while other scalar field
configurations seem to be unrealistic, including Wyman's ``other''
solution \cite{Wyman81} and its generalization \cite{Sultana15} (the 
nature of which, although not difficult to study, is shown
here for the first time). Unfortunately, the solutions discussed are 
probably unstable and, therefore, not very relevant physically 
(their stability will be studied in a future 
publication). However, any new analytical solution gives some
insight into a complicated non-linear theory and has some value. In any 
case, since the more well known FBLJNWBW
geometry has already been mapped into the general (spherical, static,
asymptotically flat) solution of Brans-Dicke theory \cite{Bronnikov,
VFAT}, it is natural to do the same for its lesser known sister, Wyman's
``other'' solution, as done in Sec.~\ref{sec:3}. The new Brans-Dicke 
solution (\ref{new1}),
(\ref{new2}) fills this gap in the literature and the special case
$\omega=0$ adds {(as shown in Sec.~\ref{sec:4})} to the catalogue of 
solutions of scale-invariant $f({\cal R})={\cal R}^2$  gravity.


\begin{acknowledgments}

This work is supported by Bishop's University and by the Natural
Sciences \& Engineering Research Council of Canada (Grant No.
2016-03803 to V.F.). A. Banijamali and B. Fazlpour would like to
thank Bishop's University for hospitality during their visit.

\end{acknowledgments}

\end{document}